# Vidange d'un réservoir

Par Thomas Gibaud [1] et Alain Gibaud, [2]
1- Université de Fribourg, département de physique, chemin du musée 3, 1700 Fribourg, Suisse
2- Université du Maine, Faculté des Sciences, UMR 6087 CNRS, 72085 Le Mans cedex 09

*Résumé*

*Nous montrons dans cet article que suivant les conditions d'écoulement un même fluide peut être considéré comme parfait ou au contraire comme visqueux. Ces propriétés sont abordées à travers l'expérience de la vidange d'un réservoir ou l'on montre que l'on passe bien d'un régime à l'autre en changeant juste la longueur du tube de sortie du réservoir vidangé. Ce changement de régime s'opère conformément à un critère que l'on défini dans la partie théorique.*

## 1. Introduction

L'objectif de cet article est de mettre, d'une part, en évidence qu'un même fluide peut très bien se comporter comme un fluide parfait ou un fluide visqueux suivant la configuration de l'expérience et d'autre part que l'on peut appréhender ce changement de comportement grâce à un nombre sans dimension. Ces propriétés sont abordées à travers l'expérience de vidange d'un réservoir par un tuyau. Cet article se décompose en trois parties. La première introduit en parallèle le modèle du l'écoulement parfait et le modèle du fluide visqueux en rapport avec l'expérience. La seconde concerne la mesure. La dernière partie concerne la discussion des résultats.

## 2. Théorie

Le système est composé d'un réservoir percé au fond d'un trou qui est prolongé par un tube. Le but de l'expérience est certes d'étudier la vidange d'un fluide mais aussi de comprendre que suivant les conditions d'écoulement un même fluide peut être considéré comme parfait ou au contraire comme visqueux. Il convient alors de déterminer un modèle pour l'écoulement visqueux et pour l'écoulement parfait ainsi que des critères simples permettant de distinguer l'un de l'autre.

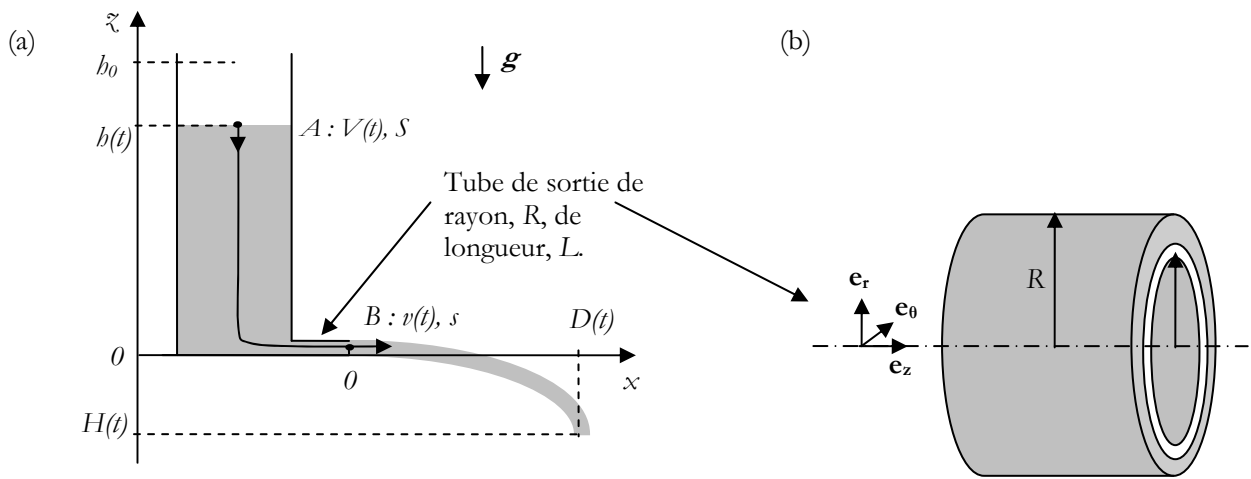

Figure 1 : (a) schéma du montage. *h* désigne la différence de hauteur entre la surface libre et le trou au fond du réservoir. $h_0$ est la hauteur initiale. La surface libre, A, est caractérisée par sa vitesse, *V*, et sa section, *S*. Le fluide s'écoule en bas par un petit tuyau, B, de longueur, *L*, de section, *s* à une vitesse, *v*. On note Φ le débit massique. (b) zoom sur le tube de sortie. Le cylindre blanc représente la couche de fluide de rayon, *r*, sur lequel on effectue le bilan des forces dans le modèle du fluide visqueux.

En supposant l'écoulement incompressible et parfait, on peut appliquer le théorème de Bernouilli ce qui permet d'obtenir la variation de *v* en fonction de *h* :

- Fluide incompressible : $\rho = cste$
- Conservation de la masse : $\Phi(A,t) = \Phi(B,t) \Leftrightarrow \rho VS = \rho vs$
- Conservation de l'énergie mécanique : $\begin{vmatrix} E(A,t) = E(B,t) \\ \text{Convention : } E_p(B) = 0 \end{vmatrix} \Leftrightarrow \frac{1}{2}\rho V^2 + \rho gh = \frac{1}{2}\rho v^2$

$$\Rightarrow v(t) \underset{s \ll S}{=} \sqrt{2gh(t)}$$

Le liquide s'écoule avec une vitesse indépendante de sa masse volumique : il n'y a pas de différence entre le mercure et l'eau ! C'est la loi de Galilée?? (Newton) sur la chute des corps transposée à l'hydrodynamique.

Si le fluide visqueux l'énergie mécanique ne se conserve plus même pour un écoulement incompressible. Afin d'obtenir la vitesse de l'écoulement à la sortie du tube, examinons les forces tangentielles s'exerçant sur une couche cylindrique de fluide de rayon $r$ de longeur $L$ (figure 1b) :

- Système invariant par translation et rotation: $\mathbf{v}(r,\theta,z) = v(r)\mathbf{e_z}$
- Force de viscosité : $f_v = \eta 2\pi r L \frac{dv}{dr}$
- Force de pression entre les 2 extémités du tube : $f_p = \pi r^2 \Delta P = \pi r^2 \rho gh$

En régime stationnaire, il y a équilibre des forces donc

$$\eta 2\pi r L \frac{dv}{dr} = \pi r^2 \rho gh$$

En intégrant cette équation et en respectant les conditions aux limites $v(R,t)=0$ on obtient la loi de Poiseuille qui relie le profil de la vitesse dans le tube à sa section:

$$v(r,t) = \frac{\rho gh(t)}{4\eta L}\left[R^2 - r^2\right]$$

L'introduction des forces de frottement dans le calcul conduise à un profil parabolique de la vitesse au sein de la conduite avec une vitesse maximale au centre et une vitesse nulle en périphérie. D'un point de vue physique on voit bien que la vitesse au centre est d'autant plus faible que le la viscosité est grande et que le tube est long. Par ailleurs elle varie linéairement avec la hauteur de chute et croît avec l'ouverture de la conduite. D'un point de vue pratique les grandeurs mesurables sont la vitesse moyenne du fluide à la sortie du tube et le débit massique. Ces deux grandeurs se déduisent de $v(r,t)$ comme suit:

$$\begin{cases} \Phi(t) = \int_0^R v(r,t)2\pi r dr = \frac{\pi R^4}{8L\eta}\rho gh(t) \\ v(t) = \frac{\Phi(t)}{s} \end{cases}$$

Les deux modèles que nous avons étudiés montrent que la vitesse instantanée d'écoulement doit évoluer en fonction de la hauteur $h(t)$ comme une loi de puissance 1/2 pour un écoulement parfait et comme une loi linéaire pour un écoulement visqueux. Ces deux comportements sont très différents et sont en principe facilement discernables. Toutefois d'autres critères peuvent être utiles pour identifier le modèle approprié. Un premier indicateur consiste à déterminer le nombre de Reynolds, noté Re. En effet, Re mesure le rapport entre les forces d'inertie et les forces de viscosité. Pour une conduite cylindrique, comme le tuyau de sortie, l'écoulement est considéré comme parfait si le nombre de Reynolds est supérieur à Re=1000 et visqueux dans le cas contraire.

$$\text{Re} = \frac{\rho v R}{\eta} \rightarrow \begin{cases} > 1000 \text{ écoulement parfait} \\ < 1000 \text{ écoulement visqueux} \end{cases}$$

La connaissance du nombre de Reynolds n'est cependant pas suffisante puisque la perte de charge dans le tuyau de sortie qui est d'autant plus grande que le tuyau est long et de faible diamètre, n'est pas prise en considération. Evaluons le rapport entre l'énergie cinétique du fluide et l'énergie dissipée par frottement dans le tube sortie :

$$\frac{E_c}{E_\eta} \approx \frac{1/2 \rho v^2}{\eta \Delta v L} \approx \text{Re}\frac{R}{L} \rightarrow \begin{cases} > 1 \text{ écoulement parfait} \\ < 1 \text{ écoulement visqueux} \end{cases}$$

Si ce rapport est grand devant 1 alors l'expérience cadre avec le modèle de l'écoulement parfait sinon il faut employé le modèle de l'écoulement visqueux. Nous constatons que le type d'écoulement dépend essentiellement pour un nombre de Reynolds donné du rapport R/L . Pour une section donnée, plus le tube est long et plus la force de frottement sera manifeste. On s'attend donc à observer le comportement visqueux pour un long tube.

## 3. Expériences

L'expérience consiste à mesurer la vitesse instantanée en fonction de la hauteur $v(h)$. La mesure de la hauteur est triviale : il suffit de graduer le réservoir. Celle de la vitesse est un peu plus délicate : 2 méthodes de mesures sont ainsi envisageables.

La première consiste à mesurer le débit, $\Phi$, en fonction du temps. Pour ce faire il suffit de déterminer le temps, $\Delta t$, nécessaire pour remplir un bécher de volume, $V_b$.

$$v = \frac{\Phi}{\Delta t} = \frac{V_b}{s \Delta t}$$

Dans cette méthode, il est important de mesurer précisément la section du tube de sortie. Une méthode simple et précise consiste à mesurer la masse du volume d'eau nécessaire pour remplir le tube. La connaissance du volume et celle de la hauteur du tube permettent de déduire la section, $s$. Afin de remplir le tube facilement on aspire l'eau à l'aide d'une poire fixée à l'extrémité du tube? Si tu aspires tu vides?. Pour éviter de laisser de l'eau dans le tube, on vide le tube très lentement avec la poire. On mesure ensuite la masse du volume ainsi récupéré. On effectue la mesure sur le plus grand tube que l'on possède, ici, L=160cm, pour obtenir la meilleure précision sur la section.

| Caractéristiques de l'expérience | $\varrho$ (kg/m³) | $\eta$ (m²/s), T=22°C | $g$ (m/s²) | $S$ (mm²) |
|---|---|---|---|---|
| | 1000 | 0.95 10⁻⁹ | 9.8 | 4654 ± 60 |
| Mesure de la Section du tube de sortie, $s$. | $L$ (cm) | Volume d'eau, $Ve$ (mL) | $s$ (mm²) | |
| | 160 ± 1 | 5.28 ± 0.10 | 3.30 ± 0.09 | |

Tableau 1 : caractéristique de l'expérience et mesure de la section du tube de sortie

La seconde méthode consiste à mesurer en fonction du temps la distance, $D(t)$, l'impact du jet à une hauteur fixée, H,. Un simple calcul balistique permet d'en déduire la vitesse, $v$ :

$$v = D\sqrt{\frac{g}{2H}}$$

Nous ne reportons pas les résultats concernant cette méthode car les résultats correspondaient systématiquement à une vitesse trop faible. Nous pensons que la cause la plus probable provient du frottement du jet par l'air.

Pour que les mesures soient pertinentes, il faut donc trouver un compromis entre le volume mesuré et le temps de mesure. Effectivement, le temps doit être suffisamment court pour que la vitesse soit une vitesse instantanée et suffisamment long pour que l'incertitude sur $\Delta t$ ne soit pas trop grande. Le même problème se pose pour le volume. Une façon astucieuse de palier au problème de la mesure dans le temps consiste à transformer le réservoir de vidange en vase de Mariotte (figure 2). On peut alors maintenir $h$ constant dans le temps, ce qui laisse tout le temps pour mesurer la vitesse, $v$. Par ailleurs, dans cette expérience $V=0$. On utilise pour les mesures cette astuce et on mesure à chaque hauteur, $h$, le temps nécessaire pour remplir une fiole graduée de 100 mL.

Pour les faibles vitesses, l'eau sort goutte à goutte du tuyau de sortie et parfois mouille et s'écoule sur l'extérieure du tuyau. Pour pallier à ce problème on a rajouté un tuyau en caoutchouc de rayon supérieur à 1 mm pour guider l'eau dans la fiole.

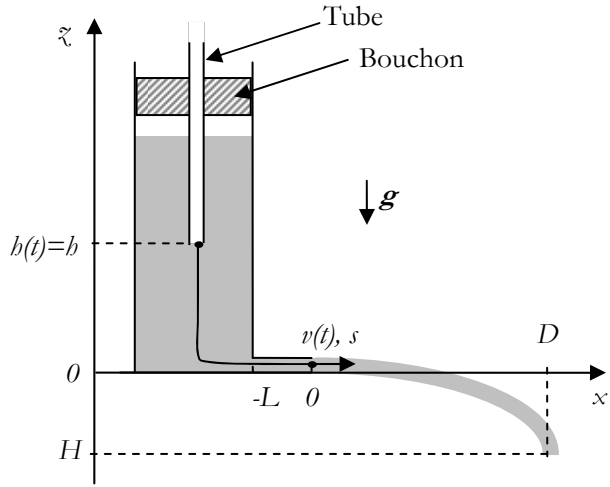

Figure 2 : Principe du vase de Mariotte. Le réservoir est hermétiquement fermé en haut par un bouchon. Le tube passant par le bouchon permet de fixer la pression de l'atmosphère non pas à la surface libre comme dans le montage précédent mais à l'extrémité inférieure du tube. Ainsi, tant que la surface libre n'est pas en dessous de la partie inférieure du tube, la hauteur pertinente pour la vidange est $h$.

## 4. Résultats et discussion

Nous avons mesuré $v$ en fonction de $h$ pour 3 longueurs différentes du tube de sortie : 5, 60 et 160 cm. Afin de comparer l'expérience au modèle de l'écoulement parfait ou visqueux, nous avons ajusté les données expérimentales avec la fonction suivante :

$$v_{fit}(A, h_{offset}, p) = A(h - h_{offset})^p$$

Cette fonction a quatre avantages : elle permet de vérifier l'ordre de la puissance, $p$, elle prend en compte un offset sur la l'estimation du zéro du tuyau de vidange, elle estime l'amplitude de la vitesse, $A$ et finalement elle ajuste aussi bien l'écoulement du fluide parfait que l'écoulement du fluide visqueux.

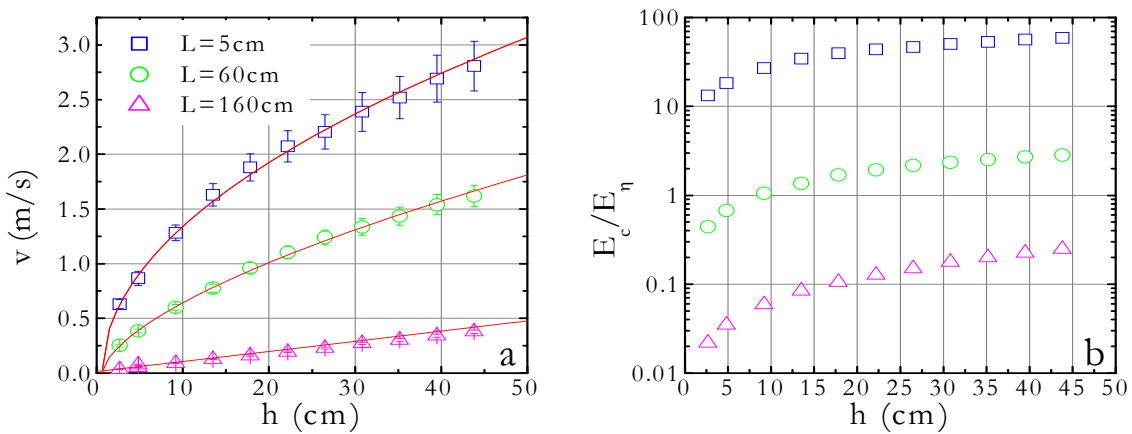

Figure 3 : (a) mesure de la vitesse en fonction de la hauteur pour différentes longueurs de tube. Les courbes en rouge représente l'ajustement (b) rapport entre l'énergie cinétique et l'énergie dissipée par frottement en fonction de la hauteur pour différente longueur de tube.

Qualitativement, on constate, se reportant à la figure 3a que plus la longueur du tube est grande, plus la vitesse est faible et que l'on passe d'un loi avec une puissance inférieure à 1 à une loi linéaire. D'après les ajustements, pour $L$=5cm, l'expérience est en parfaite adéquation avec le modèle de l'écoulement parfait alors que pour $L$=160cm, l'expérience est en parfaite adéquation avec le modèle de l'écoulement visqueux. Pour

*L*=60cm, on obtient un régime intermédiaire, ou l'exposant de la vitesse se situe entre les 2 modèles. La figure 3b représente le rapport entre l'énergie cinétique et l'énergie dissipée par frottement. On constate que la valeur de ce rapport par rapport à 1 est bien cohérente avec les modèles adoptés pour les différentes longueurs de tube de sortie. Par exemple, pour *L*=5cm ce rapport est grand devant 1 justifiant ainsi le modèle de l'écoulement parfait

| Paramètres | $A$ | $h_{offset}$ | $p$ |
|---|---|---|---|
| Expériences : | | | |
| • L=5cm | 4.36 ± 0.30 | 0.006 ± 0.006 | 0.49 ± 0.05 |
| • L=60cm | 2.81 ± 0.08 | 0.006 ± 0.003 | 0.62 ± 0.02 |
| • L=160cm | 0.89 ± 0.09 | -0.009 ± 0.010 | 0.96 ± 0.10 |
| Modèle de l'écoulement parfait | 4.42 | - | 0.5 |
| Modèle de l'écoulement visqueux : | | | |
| • L=5cm | 27.08 | - | 1 |
| • L=60cm | 2.26 | - | 1 |
| • L=160cm | 0.87 | - | 1 |

Tableau 2 : comparaison entre les résultats des ajustements et ceux de la théorie

En somme, nous avons mis en évidence que la vidange d'un réservoir permet de transformer l'énergie de pesanteur en énergie cinétique. Pour peu que l'écoulement soit parfait, ce transfert est total. Cette transformation d'énergie est mise à profit dans les barrages hydrauliques. Le jet de sortie actionne alors la rotation d'une turbine pour produire de l'électricité. Par ailleurs, en choisissant astucieusement le rapport *R/L*, on constate qu'un même fluide peut être traité comme un fluide parfait ou comme un fluide visqueux.

## Bibliographie